\documentclass[showpacs,preprintnumbers,amsmath,amssymb]{revtex4}

\usepackage{graphicx}
\usepackage{dcolumn}
\usepackage{bm}
\usepackage{color}

\begin{document}


\title{Scalar Tensor Teleparallel Dark Gravity via Noether Symmetry}

\author{Yusuf Kucukakca}%
   \email{ykucukakca@akdeniz.edu.tr}

 \affiliation{%
Department of Physics, Faculty of Science, Akdeniz University, 07058
Antalya, Turkey
   }%
\date{\today}
\begin{abstract}

In this paper, in the framework of teleparallel gravity we consider
scalar tensor theories of gravity in which scalar fields are
nonminimal coupled to torsion scalar. Noether symmetry of the
Lagrangian of such a theory for the Friedman-Robertson-Walker
spacetime is used to determine the explicit forms for the coupling
function and for the potential, and it is shown that both must be
power-law forms as a functions of the scalar field. The solutions of
the field equations for the considered models are presented by using
the results obtained from the Noether symmetry. It is shown that the
equation of state parameter in the present model can cross the
phantom divide line for a special case with the coupling function
$F(\phi) = \frac{3}{16} \phi^2$ and for the potential $V(\phi) =
\lambda \phi^{2}$.

\end{abstract}

\maketitle

\section{\label{sec:intro}Introduction}

In the past decade there is a general consensus that
today our universe is undergoing an accelerated expansion. This
result with the accelerated expansion becomes the central theme of
the modern cosmology and confirms various observational evidences
which are the observations of supernovae Type Ia  (SNe Ia)
\cite{riess99,Perll,Sperl}, cosmic microwave background radiation
(CMB) \cite{netterfield02,bennet}, and large-scale structure
\cite{tegmark04}. The standard cosmology can not clarify what causes
this cosmic acceleration and cosmologists have to seek a powerful
explanation for this observed realty in observational cosmology. In
order to explain the accelerated behaviour of the universe in
framework of General Relativity (GR) one can produce a dark fluid
with a negative pressure which named the dark energy. A simple
candidate of dark energy is the cosmological constant with the
equation of state (EoS) parameter, $w = -1$. However, the
cosmological constant model is subject to the so-called fine-tuning
and coincidence problems \cite{copeland06,durrer08}. To solve these
problems, various dynamical dark energy models have been proposed.
The simplest one is a canonical scalar field which is dubbed
quintessence dark energy that has been extensively studied in the
literature \cite{Rat,Wet,Lid,Zla}. Another one is to work a scalar
field with a negative kinetic energy, dubbed phantom dark energy
\cite{Cad,Cad2,Noj,Onn}. Finally, quintom scenario combining the
quintessence and phantom dark energy has been proposed
\cite{Li,Fen,Gao}. In such a model EoS parameter can pass below and
up the phantom divide line $w = -1$.

On the other hand alternative approach dealing with the acceleration
problem of the universe is to modified the geometric part of
Einstein field equation. Such a modifications in GR is known as
$f(R)$ gravity theory. The simplest form of this theory can be
constructed by replacing the Ricci scalar $R$ with an arbitrary
function $f(R)$ in the Einstein-Hilbert Lagrangian (for reviews see
e.g.\cite{Sot,Fel,Noj2,Cap1}).

It is known that teleparallel equivalent of GR, referred to as
teleparallel gravity, was firstly introduced by Einstein to unify
the gravity and the electromagnetism \cite{Ein1,Ein2}. In this
gravity theory, the equations of motion for any space-time are
exactly the same as of GR, but, instead of using the curvature
defined by the Levi-Civita connection, it uses the Weitzenb{\"o}ck
connection that has no curvature but only torsion. Similar to GR,
where the action is a curvature scalar, the action of teleparallel
gravity is a torsion scalar $T$. The dynamical variable of the
teleparallel gravity is the vierbein fields and the field equations
are obtained by taking variation of the action with respect to the
vierbein fields \cite{Hay}. Some modifications in the teleparallel
gravity theory have been recently proposed in order to get a
solution for the cosmic acceleration of the universe without
introducing the dark energy. A particular class of modified gravity
theories is the $f(T)$ gravity theory which can be constructed by
replacing the torsion scalar $T$ with an arbitrary function $f(T)$
in the action of the teleparallel gravity \cite{Beng,Lind,Wu,Myr}.
We note that the field equations in this theory are second-order
differential equations, while for the generalized $f(R)$ theory they
are fourth-order. Lately, $f(T)$ theory attracted much attention in
the literature, and we refer to e.g.
\cite{a1,a2,a3,a4,a5,a6,a7,a8,a9,a10,a11,a12,a13,a14,a15,a16,a17,a18}
for some relevant works.

Very recently, new scenario the so-called teleparallel dark energy
has been proposed by Geng et al in which they added the quintessence
scalar field non-minimally coupled to the torsion scalar in the
framework of teleparallel gravity \cite{geng1,geng2,Wei1,Xu1,Bani}.
They have found that such a theory has a richer structure than the
same one in the framework of general relativity. The richer
structure of non-minimally coupled scalar field with the torsion
scalar is due to exhibiting quintessence-like or phantom-like
behavior, or experiencing the phantom divide crossing in this
theory. We note that in the minimal coupling case, cosmological
model with the quintessence scalar field in teleparallel gravity is
identical to that in GR. Motivated by the teleparallel dark energy
scenario, we here extend it by replacing the coupling function term
$1+\xi \phi^2$ where $\xi$ is a non-minimal coupling constant, with
the arbitrary function $F(\phi)$. In such a scenario it is important
to determine the form of $F(\phi)$. Noether symmetry approach that
was introduced by Capozziello et al., allows one to choose the
potential and the coupling function dynamically in the scalar-tensor
gravity theory. Utilizing this approach we find the potential and
the coupling function in the teleparallel dark energy scenario and
we solve the analytically the field equations of the theory. Our
results show that the teleparallel dark energy equation of state
parameter has quintessence, phantom-like behavior or the transition
from quintessence to phantom phase in this theory. In particular,
when we choose the quadratic potential and coupling function via
Noether symmetry approach, we obtain that during the evolution of
the universe the equation of state parameter changes from
$w_{\phi}>-1$ to $w_{\phi}<-1$.

This paper is organized as follows. In the following section, the
field equations are derived from a point-like Lagrangian in a
Friedman-Robertson-Walker spacetime, which is obtained from an
action ingluding the scalar field non-minimally coupled to the
torsion scalar in the framework of teleparallel gravity. In the
Section \ref{NSA}, we search the Noether symmetry of the Lagrangian
of scalar tensor theory and give the solutions of the field
equations by using Noether symmetry approach. Finally, in the
Section \ref{conclusions}, we conclude with a brief summary of the
obtained results.

\section{Teleparallel Gravity With Scalar Field }
\label{TelGra}
 Let us consider the general action for a scalar field, non minimally coupled with the torsion scalar, in this case we
write
\begin{equation}
S=\int{ d^{4}x e \Big[F(\phi)T + \frac{1}{2}
\partial_{\mu}\phi\partial^{\mu}\phi -
V(\phi)\Big]}. \label{action2},
\end{equation}
where $e = det(e_{\mu}^i)=\sqrt{-g}$ that $e_{\mu}^i$ is tetrad
(vierbein) basis , $T$ is a Torsion scalar, $F(\phi)$ is the generic
function that describe the coupling and $V(\phi)$ is the potential
for the scalar field. Note that we use Planck units. For $F(\phi)=1+
\xi \phi^2$, the action (\ref{action2}) reduce to the action which
proposed by Geng et al \cite{geng1}. We also note that the action
(\ref{action2}) with the torsion formulation of GR is similar to the
standard scalar tensor theory where the scalar field couples to the
Ricci scalar \cite{cap2}.

The homogeneous and isotropic Friedmann-Robertson-Walker universe is
described by the metric
\begin{eqnarray}
\label{FRWmetric} ds^2 = dt^2-a^2(t)( dx^2 + dy^2 + dz^2)
\end{eqnarray}
where a(t) is the scale factor. It has been found in Ref.
\cite{ferra2} that the torsion scalar in the teleparallel gravity
can be expressed as $T = -\frac{6 \dot{a}^2}{a^2}$. Considering the
background in (\ref{FRWmetric}), it is possible to obtain a
point-like Lagrangian from action (\ref{action2})
\begin{equation}
\label{Lag} L = -6F(\phi)a \dot{a}^2 + a^3(\frac{\dot{\phi}^2}{2} -
V(\phi)),
\end{equation}
here, a dot indicates differentiation with respect to the cosmic
time $t$. It is well known that the Euler-Lagrange equations for a
dynamical system are
\begin{equation}
\label{E-L} \frac{\partial{L}}{\partial{q_{i}}} -
\frac{d}{dt}\left(\frac{\partial{L}}{\partial{\dot{q}_{i}}}\right) =
0,
\end{equation}
where $q_{i}$ are the generalized coordinates of the configuration
space. The configuration space of the Lagrangian (\ref{Lag}) is
$Q=(a,\phi)$ and whose tangent space is $TQ = (a,\phi,
\dot{a},\dot{\phi})$. Substituting Eq. (\ref{Lag}) into the
Euler-Lagrange equation (\ref{E-L}) for the scalar field, we obtain
\begin{equation}
\ddot{\phi}+3H\dot{\phi} + 6F'H^2 + V'(\phi)=0. \label{K-G}
\end{equation}
which is a Klein–Gordon equation for the coupled scalar field. From
the Euler-Lagrange equation for the scale factor $a$ by using the
Lagrangian (\ref{Lag}), we obtain the acceleration equation, namely
\begin{equation}
2\dot{H} + 3H^2=-\frac{p_{\phi}}{2F}. \label{acce}
\end{equation}
Furthermore, the energy function (Hamiltonian) associated with the
Lagrangian (\ref{Lag}) is found as
\begin{equation}
H^2=\frac{\rho_{\phi}}{6F}, \label{energy}
\end{equation}
which can be considered a constraint equation. In the above last
equations, we define the energy density $\rho_{\phi}$ and the
pressure $p_{\phi}$ of the scalar field as follows
\begin{equation}
\rho_{\phi}=\frac{1}{2}\dot{\phi}^{2} + V(\phi), \label{dens}
\end{equation}

\begin{equation}
p_{\phi}=\frac{1}{2}\dot{\phi}^{2} - V(\phi) + 4 F'H \dot{\phi}
\label{pres}.
\end{equation}
In order to solve the field equations, one has to choose a form for
the coupling function and the potential density. To do this, in the
following section we will use the Noether symmetry approach.

\section{Noether Symmetry Approach and Cosmological Solutions}
\label{NSA} 
As is well known, the Noether theorem generates a
conserved quantity in the classical mechanic. The application of
this theorem to cosmology was introduced by De Rittis et al.
\cite{rit1,rit2} and Cappoziello et al. \cite{cap3,cap4} to find
preferred solutions of the field equations and the dynamical
conserved quantity. The Noether theorem states that if the Lie
derivative of a given Lagrangian $L$ dragging along a vector field
${\bf X}$ vanishes
\begin{equation}
\pounds_{\bf X} L = 0. \label{noether}
\end{equation}
then ${\bf X}$ is a symmetry for the dynamics, and it generates the
conserved quantity. We also note that the Noether symmetry approach
allows one to choose the potential dynamically in the scalar-tensor
gravity theory \cite{rit1,rit2,cap3,cap4,san1,san2,kam,kuc1,kuc2},
and the explicit form of the function $f(R)$ of the modified $f(R)$
theories of gravity \cite{cap5,vak,ros,pal,hus,kuc3,sha}. In this
form of $f(R)$, cosmological solutions in the case of FRW metric can
describe the accelerated period of the Universe. Spherically
symmetric solutions in $f(R)$ theories of gravity have been also
found in \cite{cap6}, using the Noether symmetry approach. On the
other hand, some authors specified the form of $f(T)$ in the
modified $f(T)$ theories of gravity via Noether symmetry approach
\cite{wei2,ata,jam1,moh}.

The existence of Noether symmetry implies the existence of a vector
field ${\bf X}$ such that
\begin{equation}
{\bf X} = \alpha \frac{\partial}{\partial a} + \beta
\frac{\partial}{\partial \phi}  + \dot{\alpha}
\frac{\partial}{\partial \dot{a}} + \dot{\beta}
\frac{\partial}{\partial \dot{\phi}}  \label{vec1}
\end{equation}
where $\alpha, \beta$ and $\gamma$ are depend on $a$ and $\phi$.
Hence the condition given by (\ref{noether}) for the existence of a
symmetry gives rise to the following set of coupled differential
equations,
\begin{eqnarray}
& & \alpha + 2a \frac{\partial \alpha}{\partial a} + \frac{F'}{F}a
\beta =0, \label{neq1}
\\& & 3\alpha + 2a \frac{\partial \beta}{\partial \phi} = 0, \label{neq2}
\\& &  12 F \frac{\partial \alpha}{\partial \phi} -a^2 \frac{\partial \beta}{\partial a}  = 0,
\label{neq3} \\& &  3V\alpha + a V'\beta = 0. \label{neq4}
\end{eqnarray}
This system are obtained by imposing the fact that the coefficients
of $\dot{a}^2$, $\dot{a}\dot{\phi}$ and $\dot{\phi}^2$ vanish. Using
the separation of variable one can find the solutions of the above
set of differential equations (\ref{neq1})-(\ref{neq4}) for $\alpha,
\beta$, coupling function $F(\phi)$ and potential  $U(\phi)$ as
\begin{eqnarray}
 \alpha = -\frac{2 \alpha_{0}}{2n+3}a^{n+1} \phi^{-\frac{2n}{2n+3}}, \quad \beta =
\alpha_{0}a^{n} \phi^{\frac{3}{2n+3}}, \label{vec2}
\end{eqnarray}
\begin{equation}
F(\phi) = \frac{(2n+3)^2}{48} \phi^2,  \label{coupling}
\end{equation}
\begin{equation}
V(\phi) = \lambda \phi^{\frac{6}{2n+3}},  \label{pot}
\end{equation}
where $\alpha_{0}$, $\lambda>0$ and $n$ are a constants and
$n\neq-3/2$. If we consider the minimal coupling case i.e.
$F(\phi)=1/2$, the above Noether equations (\ref{neq1})-(\ref{neq4})
give an exponential potential. This case is completely equivalent to
standard quintessence model which has been studied in
\cite{rubano02,Demi05}. We note that for the physically acceptable
of our model, the most important requirement is $G_{eff}>0$ (Here we
define the effective gravitational constant as
$G_{eff}=\frac{1}{2F}$). Since the coupling function always positive
for all the values of $n$, this condition is provided.

In the present form the analytical solutions of the field equations
given by (\ref{K-G})-(\ref{energy}) are not straightforward to
calculate . In order to integrate our the dynamical system, we
search for a cyclic variable associated with the Noether symmetry.
If there exist such a symmetry, then one can always perform a change
of variables $(a,\phi)\rightarrow (z,u)$ such that the Lagrangian
(\ref{Lag}) becomes cyclic in one of them. We will see below that
the dynamical system involving the cyclic variable can be easily
solved. Following this procedure given by Ref.\cite{cap7}, we can
introduce the new variables as
\begin{eqnarray}
 z= \frac{2n+3}{4n \alpha_{0}}a^{-n} \phi^{\frac{2n}{2n+3}}, \quad u
 =a \phi^{\frac{2}{2n+3}}
, \label{newV}
\end{eqnarray}
in which the variable $z$ is a cyclic and $n\neq0$. Then the scale
factor and the scalar field can be rewritten as
\begin{eqnarray}
 a= u^{\frac{1}{2}}\Big(\frac{4n\alpha_{0}
 z}{2n+3}\Big)^{-\frac{1}{2n}},\nonumber \\
 \phi=u^{\frac{2n+3}{4}}\Big(\frac{4n\alpha_{0}
z}{2n+3}\Big)^{\frac{2n+3}{4n}}. \label{donap}
\end{eqnarray}
In term of these new variables, considering the coupling function
(\ref{coupling}) and  potential (\ref{pot}), the Lagrangian
(\ref{Lag}) takes a manageable form
\begin{equation}
\label{Lag2} L = (2n+3)\alpha_{0}u^{n+2}\dot{u}\dot{z} -2 \lambda
u^{3}.
\end{equation}
It is clear that the variable $z$ is cyclic for this Lagrangian. The
Euler-Lagrange equations for $z$ and $u$ and the Hamiltonian
constraint equation ($E_L = 0$) associated with this Lagrangian are,
respectively,
\begin{equation}
\label{df1}\dot{u}-\frac{\epsilon_{0}}{(2n+3)\alpha_{0}}u^{-(n+2)}=0,
\end{equation}
\begin{equation}
\label{df2} \ddot{z}+\frac{6\lambda}{(2n+3)\alpha_{0}}u^{-n}=0,
\end{equation}
\begin{equation}
\label{df3}
\dot{u}\dot{z}+\frac{2\lambda}{(2n+3)\alpha_{0}}u^{1-n}=0,
\end{equation}
where $\epsilon_{0}$ is an integration constant related with the
constant of motion. Then the solutions $z$ and $u$ are obtained as
\begin{eqnarray}\label{sol1}
 u(t) = \Big(b_{1} t +
c_{1}\Big)^{-\frac{m}{n+3}},
\end{eqnarray}
\begin{eqnarray}\label{sol2}
z(t)=b_{2}\Big(b_{1} t + c_{1}\Big)^{\frac{n+6}{n+3}}+c_{2},
\end{eqnarray}
where $c_{1}$ and $c_{2}$ are an integration constants and we also
define $b_{1}=-\frac{\epsilon_{0}(n+3)}{m\alpha_{0}(2n+3)}$ and
$b_{2}=-\frac{2 \lambda \alpha_{0} m^2
(2n+3)}{\epsilon_{0}^2(n+6)}$, $n\neq -6$. Putting the above
expressions into Eq. (\ref{donap}), we find the general solutions
for the scale factor and scalar field {\small{
\begin{equation}\label{scalef-1}
 a(t) = \Big(\frac{4\alpha_{0}b_{2}}{2n+3}\Big)^{-\frac{1}{2n}}\Big(b_{1} t +
c_{1}\Big)^{-\frac{3}{n(n+3)}},
\end{equation}}} {\small{
\begin{equation}\label{scalarf-1}
 \phi(t) =  \Big(\frac{4\alpha_{0}b_{2}}{2n+3}\Big)^{\frac{2n+3}{4n}}\Big(b_{1} t +
c_{1}\Big)^{\frac{2n+3}{2n}},
\end{equation}}}

here we assume that $c_{2}$ is zero without loss of generality. The
deceleration parameter, which is an important quantity in cosmology,
is defined by $q =-a \ddot{a}/\dot{a}^2$ , where the positive sign
of $q$ indicates the standard decelerating models whereas the
negative sign corresponds to accelerating models and $q=0$
corresponds to expansion with constant velocity.  It takes the
following form in this model
\begin{eqnarray} \label{dec1}
& &  q = -\frac{1}{3}(n^2+3n+3).
\end{eqnarray}
From Eq. (\ref{dec1}) we see that the universe is accelerating for
$n\in reel$. We can also define the equation of state parameter for
the scalar field by using Eqs. (\ref{acce})-(\ref{pres}) as
$w_{\phi}\equiv\frac{P_{\phi}}{\rho_{\phi}}=\frac{2q - 1}{3}$. Then
it can be obtained by
\begin{eqnarray} \label{eos1}
  w_{\phi} = -\frac{1}{9}(2n^2+6n+9).
\end{eqnarray}
Astrophysical data indicate that $w$ lies in a very narrow strip
close to $w = -1$. The case $w = -1$ corresponds to the cosmological
constant. For $w<-1$ the phantom phase is observed, and for $-1 < w
< -1/3$ the phase is described by quintessence. Thus, in the
intervals $-\infty < n < -3$ and $0 < n < \infty$  we have phantom
phase. If $-3 < n < 0$, then the quintessence phase occurs, where
the universe is both expanding and accelerating.
\par
Now, from the above solution we concentrate on the case $n =0, -3$
which provide, as will be shown shortly, an interesting class of
models. Firstly we consider the case $n=0$.

\subsection{Case $n= 0$} \label{case1}
For $n=0$, from Eqs. (\ref{coupling}) and (\ref{pot}) the coupling
function and the potential of scalar field result
$F(\phi)=\frac{3}{16} \phi^2$ and $V(\phi)=\lambda \phi^2$.

With the same procedure as the above section we find the new
transformations
\begin{eqnarray}
 z = -\frac{3}{2 \alpha_{0}}ln(a), \quad u
 =a^{\frac{3}{2}} \phi
, \label{newV2}
\end{eqnarray}
and the inverse transformation
\begin{eqnarray}
 a= e^{-\frac{2\alpha_{0}}{3} z}, \quad
 \phi=u e^{\alpha_{0} z}. \label{donap2}
\end{eqnarray}
We rewrite Eq. (\ref{Lag}) using the inverse transformation Eq.
(\ref{donap}) and the solutions for the coupling function $f(\phi)$
the potential $V(\phi)$ as obtained in the this section in the
following form
\begin{equation}
\label{Lag3} L = \alpha_{0}u\dot{u}\dot{z} +\frac{1}{2}\dot{u}^2-
\lambda u^{2}.
\end{equation}
This Lagrangian yields following the equations of motion
\begin{equation}
\label{df4}\dot{u}-\frac{\epsilon_{1}}{\alpha_{0}}u^{-1}=0,
\end{equation}
\begin{equation}
\label{df5} \ddot{u} + \alpha_{0}u \ddot{z} +2\lambda u=0,
\end{equation}
\begin{equation}
\label{df6} \frac{1}{2}\dot{u}^2 + \alpha_{0} u
\dot{u}\dot{z}+\lambda u^{2} + =0,
\end{equation}
where $\epsilon_{1}$ is a constant of motion for this case. The
above equations have a solutions as
\begin{equation}\label{sol3}
z(t)=-\frac{\lambda t^2}{\alpha_{0}}-\frac{c_{3}\lambda
t}{\epsilon_{1}}-\frac{\ln{\left(2 \epsilon_{1} t + \alpha_{0}
c_{3}\right)}}{4\alpha_{0}} +c_{4},
\end{equation}
\begin{eqnarray}\label{sol4}
 u(t) =  \frac{\sqrt{\alpha_{0}\left(2 \epsilon_{1} t + \alpha_{0}
 c_{3}\right)}}{\alpha_{0}},
\end{eqnarray}
where $c_{3}$ and $c_{4}$ are an integration constants. After
returning to the original variables we obtain the scale factor and
the scalar field
\begin{equation}\label{scalef-2}
 a(t) = \left(2 \epsilon_{1}t+\alpha_{0}c_{3}\right)^{\frac{1}{6}}e^{\Big[\frac{2\alpha_{0}}{3}\Big(\frac{\lambda t^2}{\alpha_{0}}+\frac{
c_{3}\lambda t}{\epsilon_{1}}-c_{4}\Big)\Big]},
\end{equation} {\small{
\begin{equation}\label{scalarf-2}
 \phi(t) =\alpha_{0}^{-1/2} \left(2 \epsilon_{1} t+\alpha_{0}c_{3}\right)^{\frac{1}{4}}e^{\Big[-\alpha_{0}\Big(\frac{\lambda t^2}{\alpha_{0}}+\frac{
c_{3}\lambda t}{\epsilon_{1}}-c_{4}\Big)\Big]}.
\end{equation}}}
Using the definition of the deceleration parameter and the equation
of state parameter one can find
\begin{eqnarray} \label{dec2}
& &  q = -1-\frac{6 \epsilon_{1}^2\left(8\lambda \epsilon_{1}^2 t^2
+8\lambda \epsilon_{1} \alpha_{0} c_{3} t +2\lambda \alpha_{0}^2
c_{3}^2 - \epsilon_{1}^2\right)}{\left(8\lambda \epsilon_{1}^2 t^2
+8\lambda \epsilon_{1} \alpha_{0} c_{3} t +2\lambda \alpha_{0}^2
c_{3}^2+ \epsilon_{1}^2\right)^2},
\end{eqnarray}
\begin{eqnarray} \label{eos2}
& &  w_{\phi} = -1-\frac{4 \epsilon_{1}^2\left(8\lambda
\epsilon_{1}^2 t^2 +8\lambda \epsilon_{1} \alpha_{0} c_{3} t +
2\lambda \alpha_{0}^2 c_{3}^2 -
\epsilon_{1}^2\right)}{\left(8\lambda \epsilon_{1}^2 t^2 +8\lambda
\epsilon_{1} \alpha_{0} c_{3} t +2\lambda \alpha_{0}^2 c_{3}^2 +
\epsilon_{1}^2\right)^2}.
\end{eqnarray}

\begin{figure}
\resizebox{0.43\textwidth}{!}{
\includegraphics{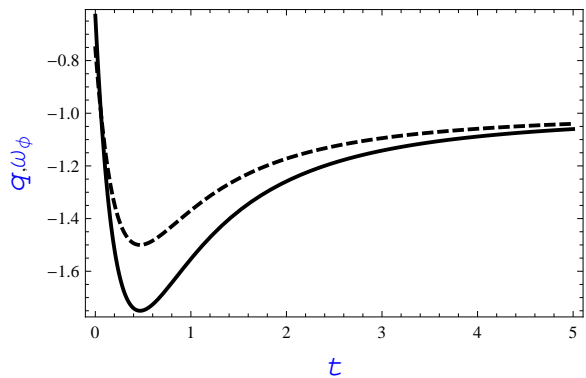}}
\caption{For $n=0$, plots of the deceleration parameter (solid line)
and the equation of state parameter (dashed line) with respect to
cosmic time $t$. We take $\epsilon_{1}=\alpha_{0}=c_{3}=1$ and
$\lambda=0.4$} \label{fig:1}
\end{figure}

In order to determine evolution of the equation of state parameter
and the deceleration parameter, we present the results in
Figure~\ref{fig:1} This figure clearly shows that crossing of the
phantom divide line $w_{\phi}=-1$ can be realized from the
quintessence phase $w_{\phi}>-1$ to the phantom phase $w_{\phi}<-1$
in our model. It is interesting to note that such a crossing is
agreement with the recent cosmological observational data
\cite{alam,ness,jass}.

\subsection{Case $n= -3$} \label{case2}
From Eqs. (\ref{coupling}) and (\ref{pot}), the coupling function
has the same form as the case $n=0$ but the potential of scalar
field results $V(\phi)=\frac{\lambda}{\phi^2}$ which is the
well-known inverse square potential and has been studied in detail
\cite{bag,agu,cope,cal}.

With the same procedure as the above section we find the new
transformations
\begin{eqnarray}
 z = \frac{a^3\phi^2}{4 \alpha_{0}}, \quad u
 =a^{-\frac{3}{2}} \phi
. \label{newV3}
\end{eqnarray}
Therefore the Lagrangian (\ref{Lag}) takes the form
\begin{equation}
\label{Lag4} L = \alpha_{0}\frac{\dot{u}\dot{z}}{u} -
\frac{\lambda}{u^{2}}.
\end{equation}
For this Lagrangian, the Euler-Lagrange equations and Hamiltonian
constrained equation provide the following equations of motion
\begin{equation}
\label{df7}\dot{u}-\frac{ \epsilon_{2}}{\alpha_{0}}u=0,
\end{equation}
\begin{equation}
\label{df8} \ddot{z} - \frac{2\lambda}{\alpha_{0}} u^{-2}=0,
\end{equation}
\begin{equation}
\label{df9}  \dot{u}\dot{z}+\frac{\lambda}{\alpha_{0}} u^{-1}=0,
\end{equation}
where $ \epsilon_{2}$ is a constant of motion for this case. The
above equations have a solutions as
\begin{eqnarray}\label{sol5}
z(t)=\frac{\lambda \alpha_{0}}{2  \epsilon_{2}^2 c_{5}^2}
e^{-\frac{2  \epsilon_{2} t}{\alpha_{0}}} +c_{6},
\end{eqnarray}
and
\begin{eqnarray}\label{sol6}
 u(t) =  c_{5} e^{\frac{  \epsilon_{2} t}{\alpha_{0}}},
\end{eqnarray}
where $c_{5}$ and $c_{6}$ are integration constants. After returning
to the original variables we obtain the scale factor and the scalar
field, respectively,
\begin{eqnarray}\label{scalef-3}
 a(t) = \left[\frac{2\alpha_{0} e^{-\frac{2  \epsilon_{2} t}{\alpha_{0}}}}{ \epsilon_{2}^2 c_{5}^4}
 \left(\lambda \alpha_{0} e^{-\frac{2  \epsilon_{2} t}{\alpha_{0}}}+2  \epsilon_{2}^2 c_{5}^2 c_{6}\right)\right]^{1/6},
\end{eqnarray}
and
\begin{eqnarray}\label{scalarf-3}
 \phi(t) =\left[\frac{2\alpha_{0} e^{\frac{2  \epsilon_{2} t}{\alpha_{0}}}}{ \epsilon_{2}^2 }
 \left(\lambda \alpha_{0} e^{-\frac{2  \epsilon_{2} t}{\alpha_{0}}}+2  \epsilon_{2}^2 c_{5}^2 c_{6}\right)\right]^{1/4}.
\end{eqnarray}

\begin{figure}
\resizebox{0.44\textwidth}{!}{
\includegraphics{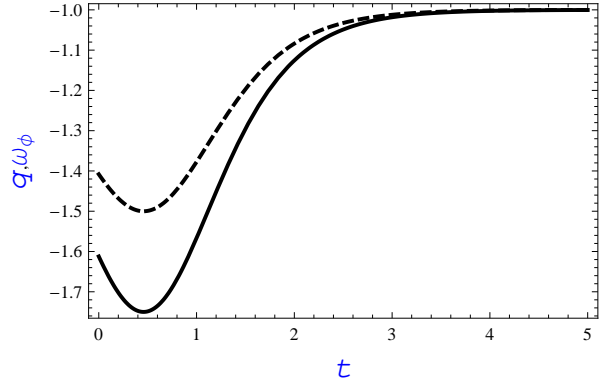}}
\caption{For $n=-3$, plots of the deceleration parameter (solid
line) and the equation of state parameter (dashed line) with respect
to cosmic time $t$. we choose $\lambda=\alpha_{0}=c_{5}=c_{6}= 1$
and $\epsilon_{2}=-1$.} \label{fig:2}
\end{figure}

\begin{figure}
\resizebox{0.44\textwidth}{!}{\includegraphics{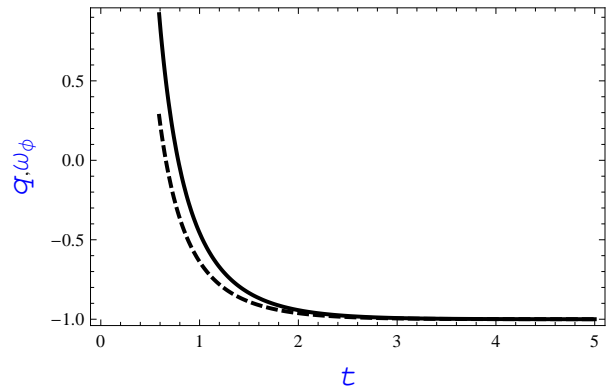}}
\caption{For $n=-3$, plots of the deceleration parameter (solid
line) and the equation of state parameter (dashed line) with respect
to cosmic time $t$. we choose $\lambda=\alpha_{0}=c_{5}=1,c_{6}= -1$
and $\epsilon_{2}=-1$.} \label{fig:3}
\end{figure}

In this case the deceleration and equation of state parameters
become

\begin{eqnarray} \label{dec2}
& &  q = -1-\frac{3\lambda \alpha_{0}  \epsilon_{2}^2 c_{5}^2 c_{6}
e^{-\frac{2  \epsilon_{2} t}{\alpha_{0}}}}{\left(\lambda \alpha_{0}
e^{-\frac{2  \epsilon_{2} t}{\alpha_{0}}}+ \epsilon_{2}^2 c_{5}^2
c_{6}\right)^{2}},
\end{eqnarray}
\begin{eqnarray} \label{eos2}
& &  w_{\phi} = -1-\frac{2\lambda \alpha_{0}  \epsilon_{2}^2 c_{5}^2
c_{6} e^{-\frac{2  \epsilon_{2} t}{\alpha_{0}}}}{\left(\lambda
\alpha_{0} e^{-\frac{2  \epsilon_{2} t}{\alpha_{0}}} +
\epsilon_{2}^2 c_{5}^2 c_{6}\right)^{2}}.
\end{eqnarray}
If we choose $c_{6} = 0$, we obtain the de Sitter solution. In
Figure~\ref{fig:2} and Figure~\ref{fig:3}  behaviors of these
parameters, by a suitable choice of the constants $b_{6}$ or
$\alpha_{0}$, are depicted in terms of the cosmic time $t$ . From
these figures we have both the phantom ($b_{6}>0$ or
$\alpha_{0}>0$), and quintessence phase ($b_{6}<0$ or
$\alpha_{0}<0$). In this case, the phantom divide line is never
crossed.

\section{Conclusions} \label{conclusions}

It has recently been proposed as an alternative dark energy scenario
\cite{geng1,geng2,Wei1,Xu1,Bani} and so-called teleparallel dark
energy. In these literatures,authors have added a scalar field to
the Teleparallel Equivalent to General Relativity, allowing for a
non minimal coupling between the field and the torsion scalar.
Motivated by this scenario, in the present work, we have
investigated the scalar tensor dark energy models, where the
gravitational part of the Lagrangian is $F(\phi)T$. Such a model may
have a richer structure than the same one in the framework of GR. If
we choose the coupling function as $F(\phi)=1+\xi\phi^2$, then our
model reduced to Lagrangian which proposed by Geng et al
\cite{geng1}. As we said above, the Noether symmetry approach a very
important tool, because it guarantees the conservation laws and
restricts the possible expressions for the coupling function and for
the potential of scalar field in the framework of teleparallel
gravity. Further, this approach leads to a variable transformation
that usually makes it possible to obtain exact and general solutions
for the evolution of the scale factor and the scalar field. Here,
using the Noether symmetry approach we construct the explicit forms
of the coupling function and potential as $F(\phi) =
\frac{(2n+3)^2}{48} \phi^2$ and $V(\phi) = \lambda
\phi^{\frac{6}{2n+3}}$. Such a models yields a power law expansion
for the cosmological scale factor (see Eq. (\ref{scalef-1})). We
have also presented the equation of state parameter for our model.
It has been turned out that a phantom like dark energy for the
intervals $-\infty < n < -3$ and $0 < n < \infty$ and a quintessence
like dark energy for the interval $-3 < n < 0$ occur. In two special
cases, for $n=0$, we have obtained the coupling function as $F(\phi)
= \frac{3}{16} \phi^2$ and the potential as $V(\phi) = \lambda
\phi^{2}$. For the model, we present the results in Fig.~1 where
shows that crossing of the phantom divide line $w_{\phi}=-1$ can be
realized from the quintessence phase $w_{\phi}>-1$ to the phantom
phase $w_{\phi}<-1$. For $n=-3$, the coupling function has a same
form but the potential has inverse square form. Our results for the
equation of state parameter have depicted in Fig.~2 and Fig.~3. From
these figures we observed that the phantom divide line is never
crossed. Therefore, by considering the above results it seems that
the quadratic potential $V(\phi) = \lambda \phi^{2}$ is a better
choice in studying the phantom divide crossing cosmology in the
context of nonminimal scalar tensor teleparallel dark energy.

\begin{acknowledgments}
This work was supported by Akdeniz University, Scientific Research
Projects Unit.
\end{acknowledgments}
\newpage

\end{document}